\documentstyle[twoside,fleqn,npb,epsfig]{article}
\newcommand{\AmS}{{\protect\the\textfont2
  A\kern-.1667eMlower.5ex\hbox{M}\kern-.125emS}}

\def\Journal#1#2#3#4{{#1} {\bf #2}, #3 (#4)}

\def\NPB{{\em Nucl. Phys.} B}

\def\ZPC{{\em Z. Phys.} C}

\title{QCD at the Tevatron: Jets and Fragmentation}

\author{V.\ Daniel Elvira\address{Fermi National Accelerator Laboratory,
        P.O.Box 500, Batavia, IL 60510}
                \thanks{Representing the D\O\ Collaboration}}
       
\begin{document}

\begin{abstract}
At the Fermilab Tevatron energies,
($\sqrt{s}$=1800~GeV and $\sqrt{s}$=630~GeV), jet production is the dominant
process. During the period 1992-1996, the D\O\ and CDF experiments
accumulated almost 100~pb$^{-1}$ of data and performed the most
accurate jet production measurements up to this 
date. These measurements and the NLO-QCD 
theoretical predictions calculated during the last decade, have 
improved our understanding of QCD, our knowledge of the proton 
structure, and pushed the limit to the scale associated with quark 
compositeness to 2.4-2.7~TeV. In this paper, we present the most 
recent published and preliminary measurements on jet production and 
fragmentation by the D\O\ and CDF collaborations.
\end{abstract}

\maketitle

\section{Introduction}

Quantum chromodynamics (QCD) describes the inelastic scattering 
between a proton and an antiproton as a hard collision between 
their constituents or partons: quarks or gluons. After the 
collision, the outgoing partons hadronize into streams of particles called
jets. At the Tevatron energies,
($\sqrt{s}$=1800~GeV and $\sqrt{s}$=630~GeV), jet production is the dominant
process. During the period 1992-1996, the D\O\ and CDF experiments
accumulated almost 100~pb$^{-1}$ of data and performed the most
accurate jet production measurements up to this 
date. 
Among the results published in that period and subsequent years, we can cite
inclusive jet cross sections, dijet angular distributions, and dijet 
mass cross sections~\cite{incd0,inccdf,angd0,angcdf,dijetd0,dicdf,bfkld0}.   
At the same time, predictions for jet production rates have 
improved in the early nineties with next-to-leading order (NLO) perturbative
QCD calculations~\cite{theory} and more accurate
parton distribution functions (pdf)~\cite{pdfs}.
The high center-of-mass energy at the Tevatron and the unprecedented
accuracy of the measurements, together with the NLO-QCD theoretical
predictions derived during the last decade, have improved our understanding
of QCD, our knowledge of the proton structure, and pushed the limit to the
scale associated with quark compositeness.

In this paper, we include a summary of some of the most significant jet
results published by~CDF and D\O\, as well as their most recent
preliminary measurements. Jet cross sections in forward pseudorapidity
regions, cross sections of dijets separated by large pseudorapidity
intervals, and subjet and particle multiplicity measurements provide
information on parton distribution functions, probe BFKL dynamics, explore
the jet structure, and study the hadronization process.
In the first sections, we describe how jets are selected, reconstructed and
calibrated at D\O\ and CDF. They are followed by sections on each measurement
and the conclusion.

\section{Jet Reconstruction and Data Selection}

For most of the analyses presented here jets are reconstructed using an 
iterative fixed cone algorithm
with a cone radius of $\cal{R}$=0.7 in $\eta$--$\phi$ 
space~\cite{d0the}, 
(pseudorapidity is defined as 
$\eta = -{\rm ln}[{\rm tan}\frac{\theta}{2}]$). This 
algorithm is applied to calorimeter towers without making use of 
tracking information,
except for the determination of the interaction vertex.  
The D\O\ subjet multiplicity measurement uses a $K_{T}$ 
algorithm \cite{ES,cat93,cat93} on 
calorimeter towers, with
a resolution parameter $\cal{D}$=1 (see Ref.~\cite{rob99}.)
The CDF particle multiplicity results use fixed cone algorithms with
different cone sizes, based on particle information at the tracking level.

The offline data selection procedure eliminates background caused by
electrons, photons, noise, or cosmic rays. In the case of D\O\, it follows 
the methods described in Refs.~\cite{levan,krane}. 

\section{Energy Corrections}

The jet energy scale correction, described by D\O\ 
in Ref.~\cite{escale}, removes
instrumentation effects associated with calorimeter response, showering, and
noise, as well as the contribution from spectator partons (underlying event).

The D\O\ energy scale correction corrects the jet $E_{T}$ from their 
reconstructed value to their ``true'' $E_{T}$ on average (energy
of a jet defined from final state hadrons). An unsmearing correction is
applied later to remove the effect of a finite $E_{T}$ 
resolution~\cite{levan}. CDF corrects both for scale and resolutions using
a Monte Carlo simulation tuned to represent the data.

\section{The Inclusive Jet Cross Section at $\sqrt{s}$~=~1800~GeV}

The inclusive jet cross section is measured by both the
D\O\ (in $|\eta| < 0.5$ and $0.1 < |\eta| < 0.7$) and the CDF
experiments (in $0.1 < |\eta| < 0.7$). 
It is defined as:
\begin{equation}
\frac{d^{2}\sigma}{dE_{T} d\eta} = 
\frac{N_{i}}{{\cal{L}}_{i} \epsilon_{i} \delta E_{T} \Delta \eta}
\label{eq:inc_cross_section}
\end{equation}
 where $N_i$ is the number of accepted jets in $E_{T}$ bin $i$ of width
 $\Delta E_{T}$, ${\cal
 L}_{i}$ is the integrated luminosity, $\epsilon_{i}$ is the
 efficiency of the trigger, vertex selection, and the jet quality
 cuts, and $\Delta \eta$ is the width of the pseudorapidity bin.

\begin{figure}[htbp]
\begin{minipage}[t]{3.05in}
\vskip-1cm
\centerline{\psfig{figure=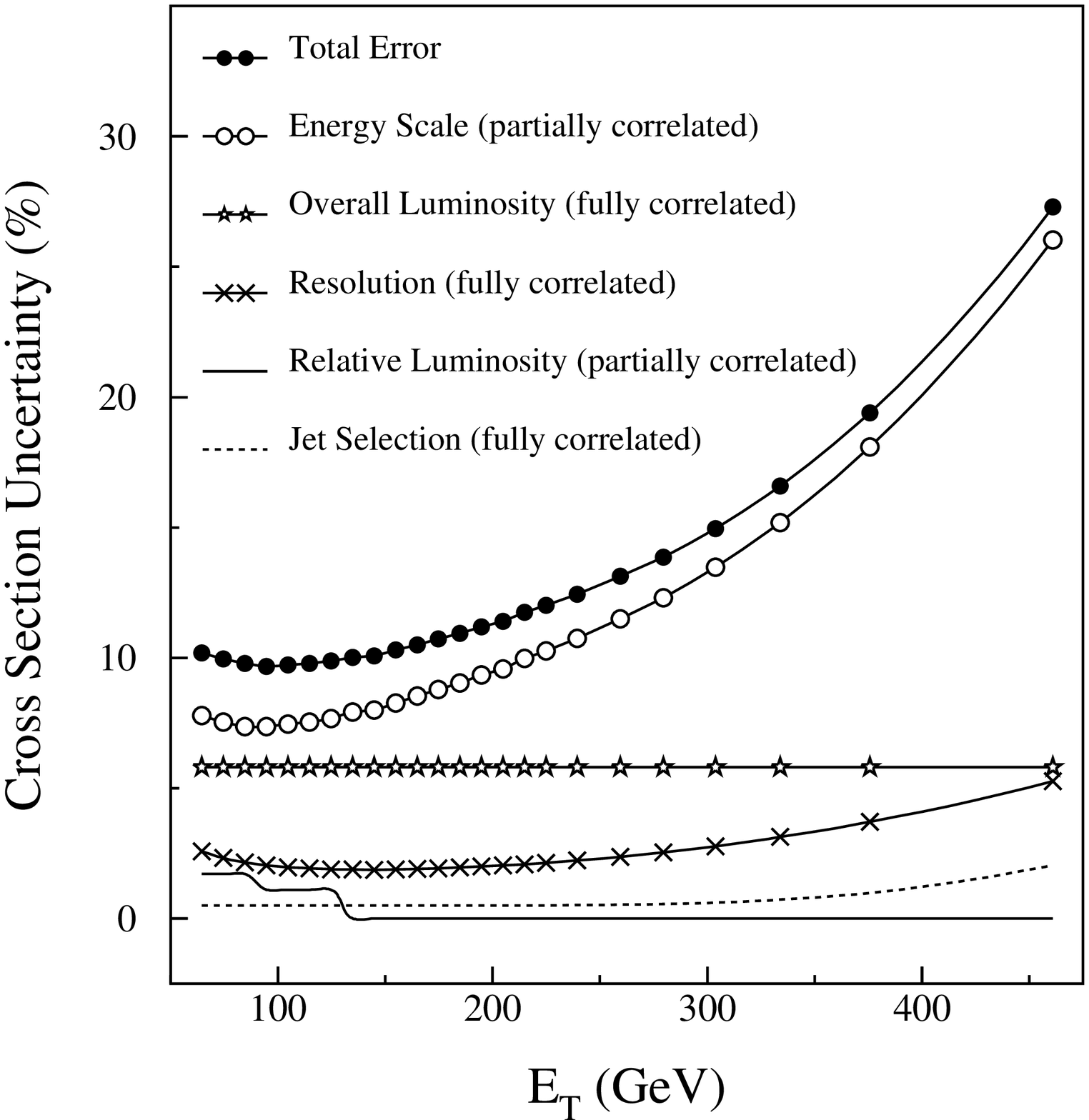,width=3.0in}}
\vskip-0.5cm
\caption{Contributions to the D\O\ jet ($ |\eta| \leq 0.5$)
    cross section uncertainty plotted by component.}
\label{fig:error_components}
\end{minipage}
\end{figure}

 Figure~\ref{fig:error_components}
 shows the various uncertainties for the D\O\ ($|\eta| < 0.5$) cross
 section.  The second outermost curve shows the error on the
 energy scale which varies from $8\%$ at low $E_{T}$ to $30\%$ at 450~GeV 
 and dominates the total error.  
 Most of the systematic uncertainties of the inclusive jet cross
 section are highly correlated as a function of $E_{T}$. 

\begin{figure}[t]
\begin{minipage}[t]{3.05in}
\centerline{\psfig{figure=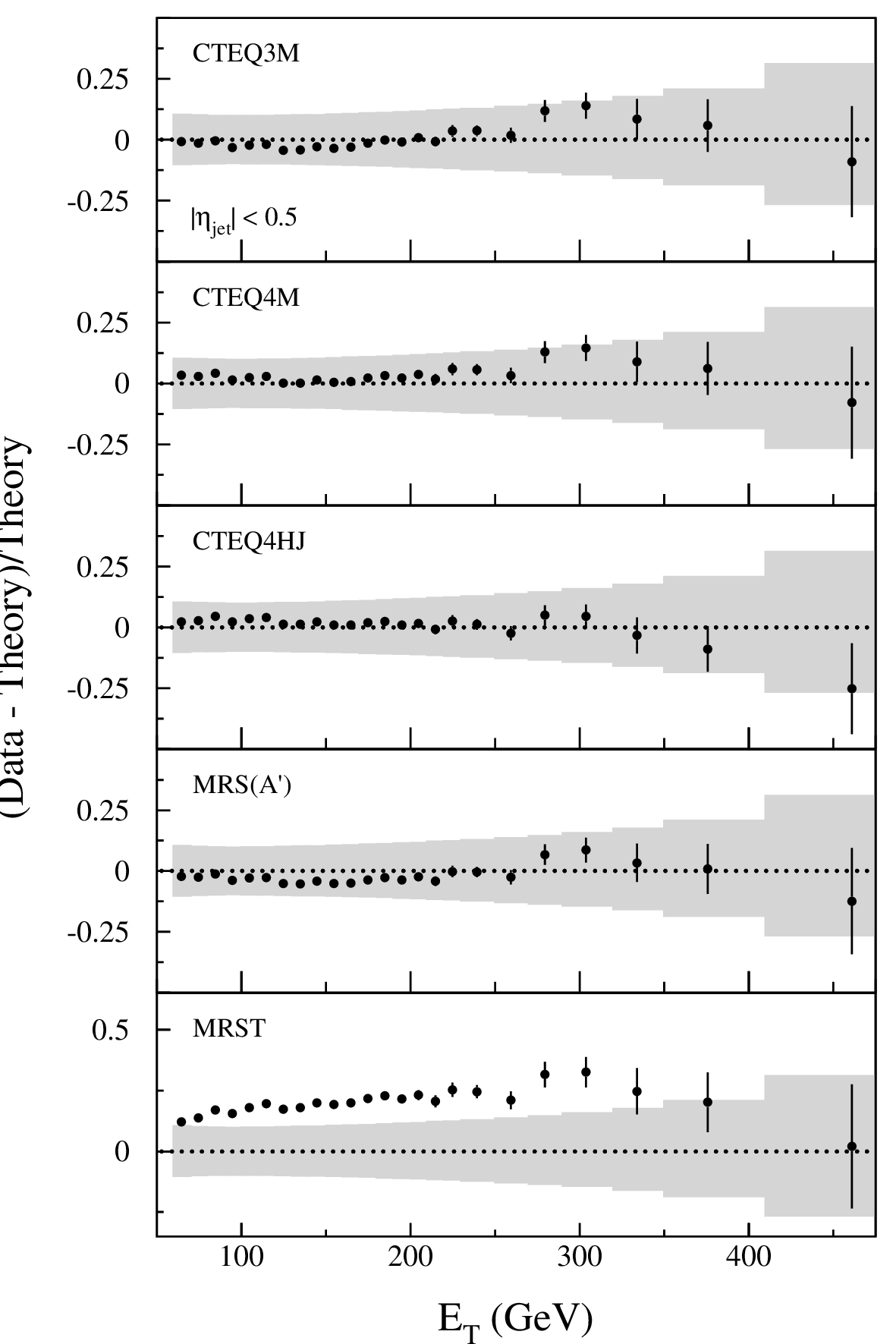,width=3.0in}}
\vskip-1cm
\caption{The difference between D\O\ data and {\sc jetrad} QCD predictions
normalized to predictions for $|\eta| < 0.5$.  The shaded region
represents the $\pm1\sigma$ systematic uncertainties about the
prediction.}
\vskip-1cm
\label{fig:data_cteq3M}
\end{minipage}
\end{figure}

 Figure~\ref{fig:data_cteq3M} show the
 fractional difference between the data, $D$, and a {\sc jetrad}
 theoretical prediction, $T$, normalized by the prediction,
 ($(D-T)/T$), for $|\eta| < 0.5$.
 The {\sc jetrad} prediction was generated with
 $\mu=0.5E_{Tmax}$, ${\cal{R}}_{\rm sep}=1.3$ and several different
 choices of pdf.  The error bars represent statistical errors only.
 The outer bands represent the total cross section error excluding the
 $5.0\%$ luminosity uncertainty. Given the experimental and
 theoretical uncertainties, the predictions are in agreement with the
 data; in particular, the data above $E_{T} = 350$~GeV show no
 indication of an excess relative to QCD.

 The data and theory can be compared quantitatively with a $\chi^{2}$
 test incorporating the uncertainty covariance matrix. The
 $\chi^{2}$ is given by:
\begin{equation}
\chi^{2} = \sum_{i,j} \delta_{i} V_{ij}^{-1} \delta_{j}
\end{equation}
 where $\displaystyle{\delta_{i}}$ is the difference between the data
 and theory for a given $E_{T}$ bin, and $\displaystyle{V_{ij}}$ is
 element $i,j$ of the covariance matrix:
\begin{equation}
V_{ij} = \rho_{ij} \cdot \Delta \sigma_{i} \cdot \Delta \sigma_{j}.
\end{equation} 
 where $\displaystyle{\Delta \sigma}$ is the sum of the systematic
 error and the statistical error added in quadrature if $i=j$ and the
 systematic error if $i \neq j$, and $\displaystyle{\rho_{ij}}$ is the
 correlation between the systematic uncertainties of $E_{T}$ bins.  

\begin{figure}[t]
\begin{minipage}[t]{3.05in}
\centerline{\psfig{figure=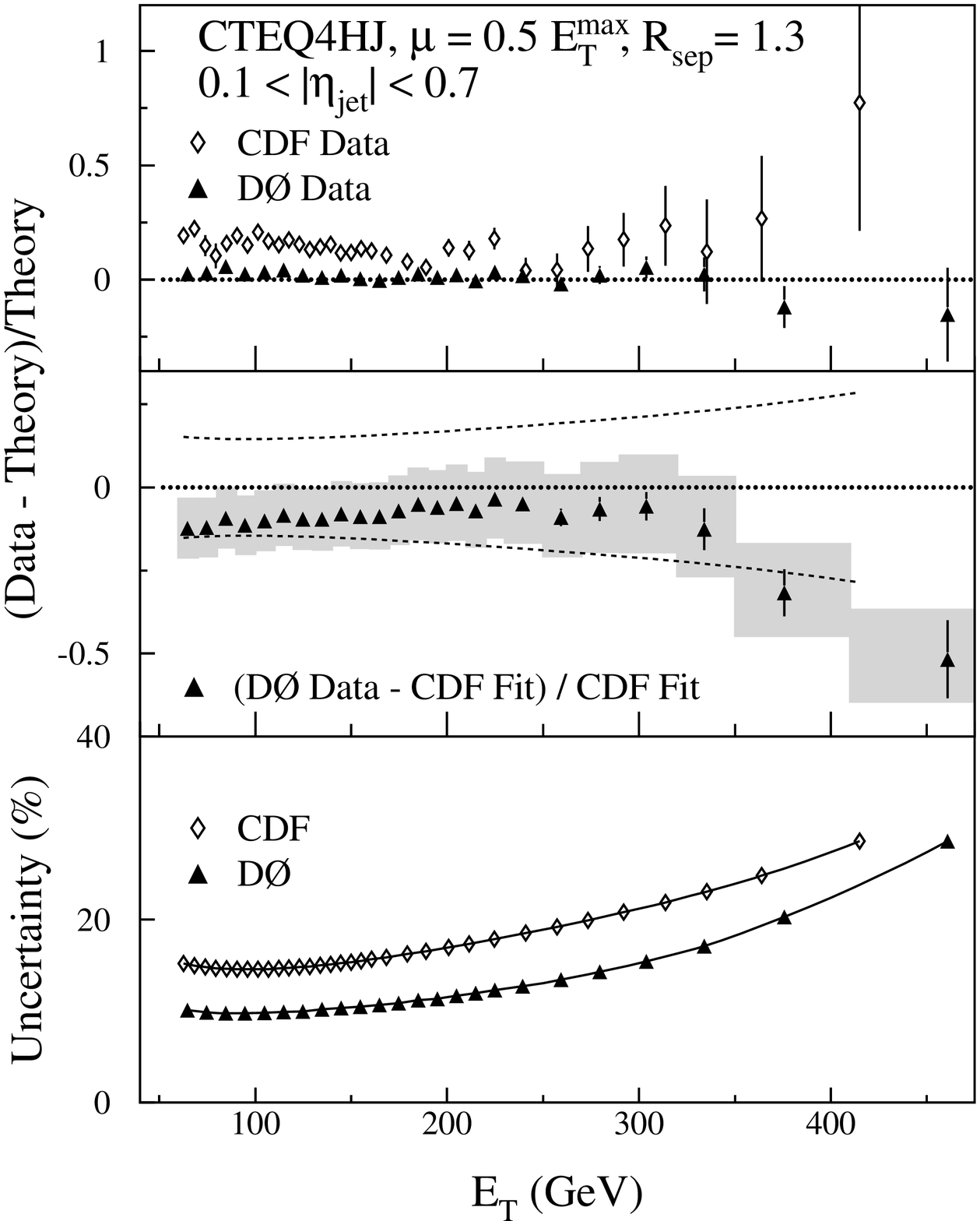,width=3.0in}}
\vskip-1cm
  \caption{Top: Normalized comparisons of the D\O\ data and of the 
  CDF data to a {\sc jetrad} prediction (CTEQ4HJ and
  $\mu = 0.5 E_{Tmax}$).  Middle: Difference between the data and
  smoothed results of CDF normalized to the latter.  The shaded region
  represents the $\pm1\sigma$ systematic uncertainties about the D\O\
  data. The dashed curves show the $\pm1\sigma$ systematic
  uncertainties about the smoothed CDF data. Bottom: A comparison of
  the systematic uncertainties of both experiments.}  
\vskip-0.5cm
\label{inclusive:Fig_cdf}
\end{minipage}
\end{figure}

 All but one of the {\sc jetrad} predictions
 adequately describe the
 $|\eta| \leq 0.5$ and $0.1 \leq |\eta| \leq 0.7$ (shown
 in Figure~\ref{inclusive:Fig_cdf}) cross sections
 (probabilities for $\chi^{2}$ to exceed the calculated value are
 between $10\%$ and $86\%$). The prediction using CTEQ4HJ and
 $\mu = 0.5 E$ produces the highest probability for both
 measurements.  The prediction with the MRSTGD pdf has a probability
 of agreement with the data of $0.3\%$, and is incompatible with the
 data.

 The top panel in Fig.~\ref{inclusive:Fig_cdf} shows $(D-T)/T$ 
 for the D\O\
 data in the $0.1 \leq |\eta| \leq 0.7$ region relative to a {\sc
 jetrad} calculation using the CTEQ4HJ pdf, $\mu=0.5E_{Tmax}$, and
 ${\cal{R}}_{\rm{sep}}=2.0{\cal{R}}$. Also included is the
 published CDF measurement from the 1992-93 Tevatron running
 period~\cite{inccdf} relative to the same {\sc jetrad} prediction.
 The CDF measurement shows an excess with respect to the theory
 at high $E_{T}$, which can be accommodated by adjusting the gluon
 pdf (CTEQ4HJ set). 
 If we include
 the systematic uncertainties of the two experiments (CDF's
 uncertainties in Ref.~\cite{inccdf} ) in a
 covariance matrix, the $\chi^{2}$ is 30.8 for 24 degrees
 of freedom (probability of
 $16\%$), representing acceptable agreement between D\O\ and CDF.

\section{Inclusive Jet Cross Section at $\sqrt{s}$= 630~GeV}

\begin{figure}[t]
\vskip-1cm
\begin{minipage}[t]{3.05in}
\centerline{\psfig{figure=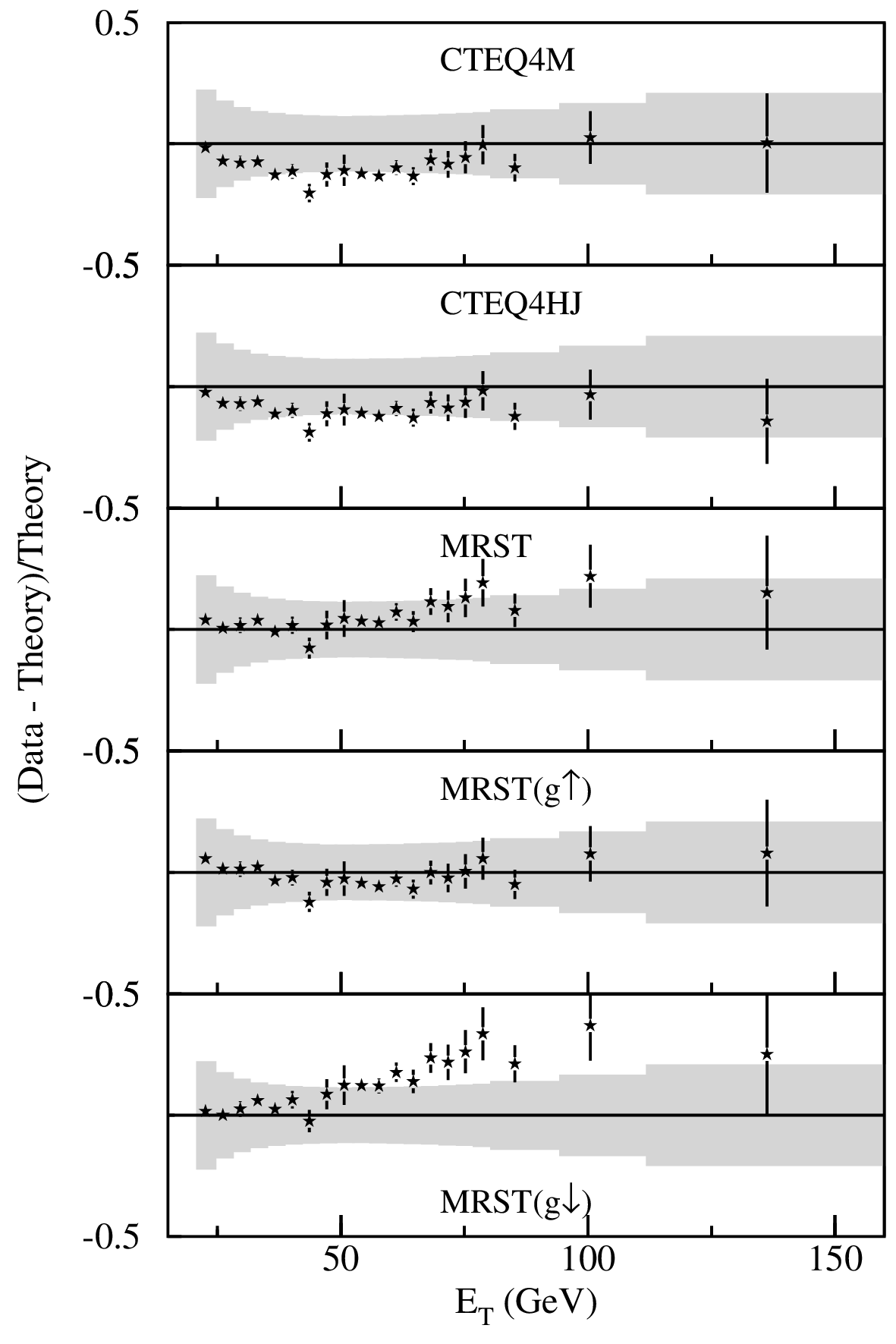,width=3.0in}}
\vskip-1cm
  \caption{(D-T)/T for the D\O\ 630~GeV jet cross section.
  ($|\eta_{jet}| < 0.5$). The solid
  stars represent the data compared to the calculation for $\mu = 0.5
  E_{Tmax}$ and the pdfs CTEQ4M , CTEQ4HJ , MRST , MRSTGU and
  MRSTGD . The shaded region represents the $\pm1\sigma$ systematic
  uncertainty.}
\vskip-1cm
\label{fig:dtt_4HJ_et2_2}
\end{minipage}
\end{figure}

 Figure~\ref{fig:dtt_4HJ_et2_2} shows the 
 fractional
 difference between the D\O\ data and several {\sc jetrad} predictions
 given different choices of renormalization scale and pdf. These NLO
 QCD predictions are in reasonable agreement with the data.  The data
 and predictions are compared quantitatively with a $\chi^{2}$ test.
 All but two of the {\sc jetrad} 
 predictions adequately describe the cross section at 
 $\sqrt{s} = 630$~GeV  (the probabilities
 for $\chi^{2}$ to exceed the calculated values are between $10\%$ and
 $74\%$). The prediction using MRSTGU and $\mu = 0.5 E_{Tmax}$
 produces the highest probability. The prediction with MRSTGD pdf
 and $\mu = 0.5 E_{Tmax}$, and CTEQ3M pdf and $\mu = 2E_{Tmax}$ are
 ruled out by the D\O\ measurement (agreement probability $\le 0.4\%$).

\section{The Ratio of Jet Cross Sections}

 The dimensionless inclusive jet cross section
 is given by
\begin{equation}
\sigma_{{\rm CM}} = \frac{E_{T}^{3}}{2\pi} \frac{d^{2}\sigma}{dE_{T} d\eta}, 
\label{eq:eq_r}
\end{equation}
 where $\displaystyle{{d^{2}\sigma}/{dE_{T} d\eta}}$ is given by
 Eq.~\ref{eq:inc_cross_section}, and $x$ is the center-of-mass
 energy. 
 A naive parton model with no $Q^{2}$ dependence of the pdfs, and therefore
no running of $\alpha_{s}$, predicts this cross section ratio to be unity, 
that is independent of the center-of-mass energy of the 
$\overline{p}p$ system. The observable is nearly insensitive to the choice of parton
 distribution functions. It is, therefore, a
 more stringent test of QCD matrix elements.   

D\O\ has measured the ratio of inclusive jet cross sections in the
central pseudorapidity bin ($|\eta|<$0.5)~\cite{ratiod0}. 
This quantity is calculated in
 bins of identical $x_T$:
\begin{equation}
R\left(x_T \right) = 
\frac{\sigma_{630}\left(x_T \right)}{\sigma_{1800}\left(x_T \right) }.
\end{equation}
The inclusive jet cross
 section errors are highly correlated as a function of $E_{T}$ and
 center-of-mass energy and will cancel in the ratio. The energy
 scale uncertainty dominates the total error in the ratio too.

\begin{figure}[htbp]
\vskip-0.5cm
\begin{minipage}[t]{3.05in}
\centerline{\psfig{figure=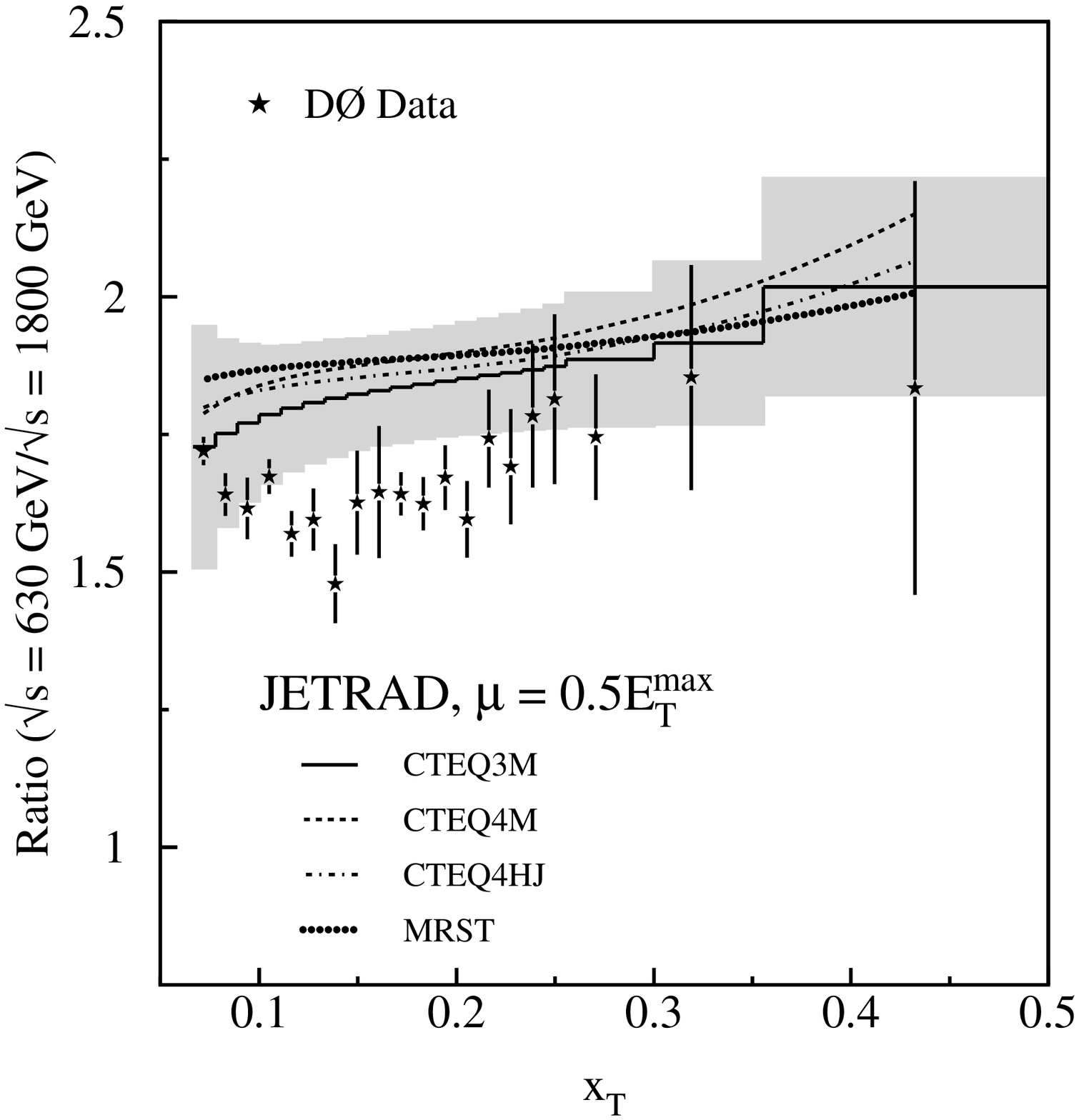,width=3.0in}}
\vskip-1cm
\hskip1cm
 \caption{The D\O\ ratio of dimensionless cross sections 
 compared with {\sc
 jetrad} predictions ($\mu = 0.5 E_{Tmax}$ and the CTEQ3M ,
 CTEQ4M , CTEQ4HJ , or MRST pdfs). The shaded band represents
the $\pm 1\sigma$ systematic uncertainty.}
\label{fig:rat_pdf}
\end{minipage}
\vskip-0.5cm
\end{figure}

 Figure~\ref{fig:rat_pdf} shows the ratios of cross
 sections with {\sc jetrad} predictions using different pdfs. 
 The measured ratios lie approximately
 $10\%$ below the theoretical predictions, which have an uncertainty
 of about $10\%$. The $\chi^{2}$ values lie
 in the range 15.1--24 for 20 degrees of freedom (corresponding to
 probabilities in the range $28\%$ to $77\%$). The best agreement
 occurs for extreme choices of renormalization scales ($\mu = 0.25,
 2 E_{Tmax}$). 

 In general, the 
 NLO-QCD predictions yield satisfactory agreement with the D\O\ data 
 for standard choices of renormalization scale or pdfs. In terms of the
 normalization, however, the absolute values of the standard
 predictions lie consistently and significantly higher than the data.

\begin{figure}[htbp]
\vskip-1cm
\begin{minipage}[t]{3.05in}
\centerline{\psfig{figure=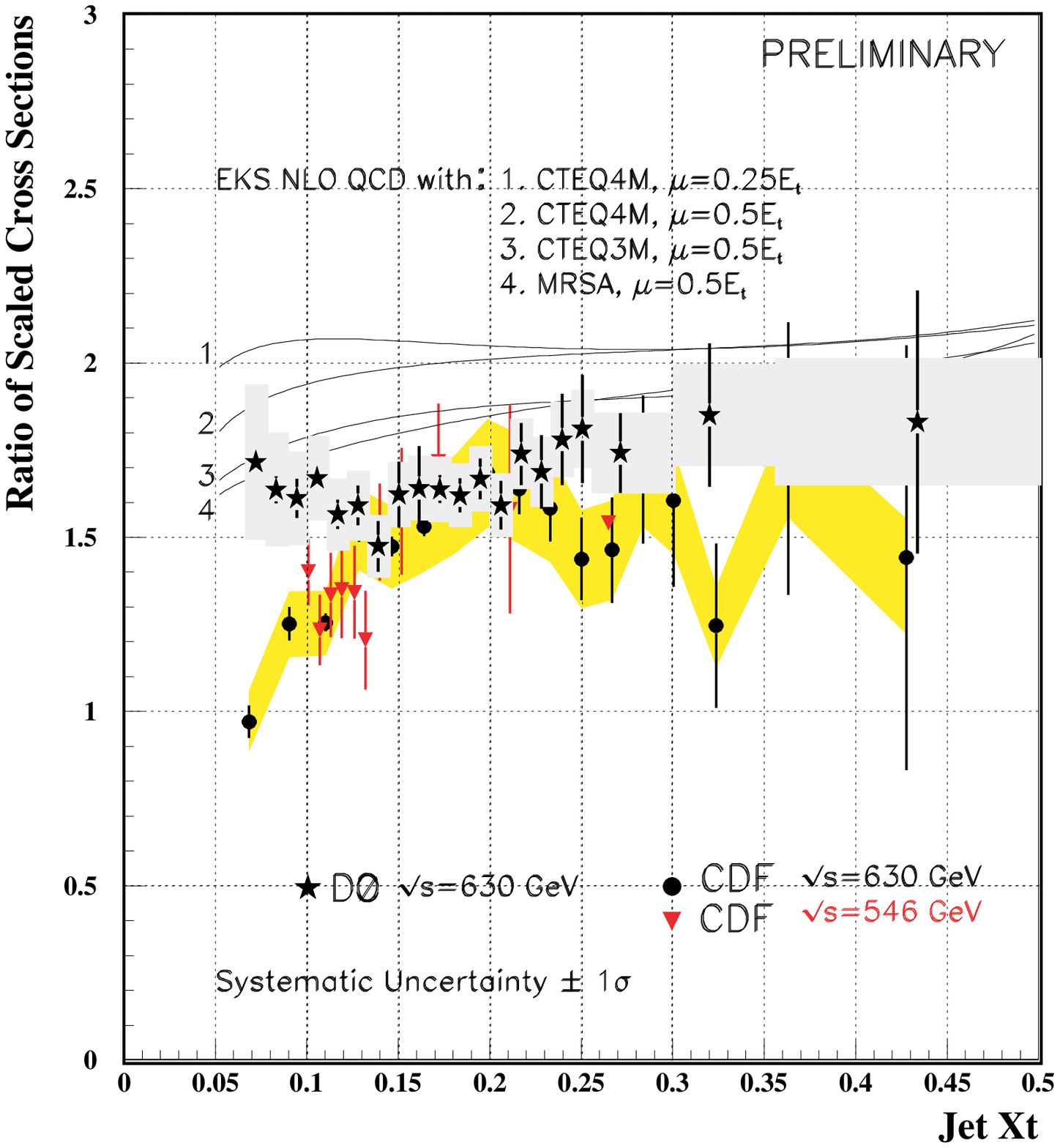,width=3.0in}}
\vskip-1cm
  \caption{The D\O\ and CDF ratio of dimensionless cross sections compared 
  with each other and the {\sc jetrad} predictions.}
\label{fig:cdfratio}
\end{minipage}
\end{figure}

 CDF has also measured the ratio of central jet cross sections. 
 Figure~\ref{fig:cdfratio} shows a preliminary result by CDF compared with
 NLO-QCD predictions (and the D\O\ measurement). Although the data and 
 the theory agree in shape at
 high $x_{T}$, there is a significant deviation at low values.

\section{The Forward Jet Cross Sections at $\sqrt{s}$~=~1800~GeV}

\begin{figure}[htb]
\vskip0.5cm
\hskip-0.5cm
\begin{minipage}[t]{2.9in}
\centerline{\psfig{figure=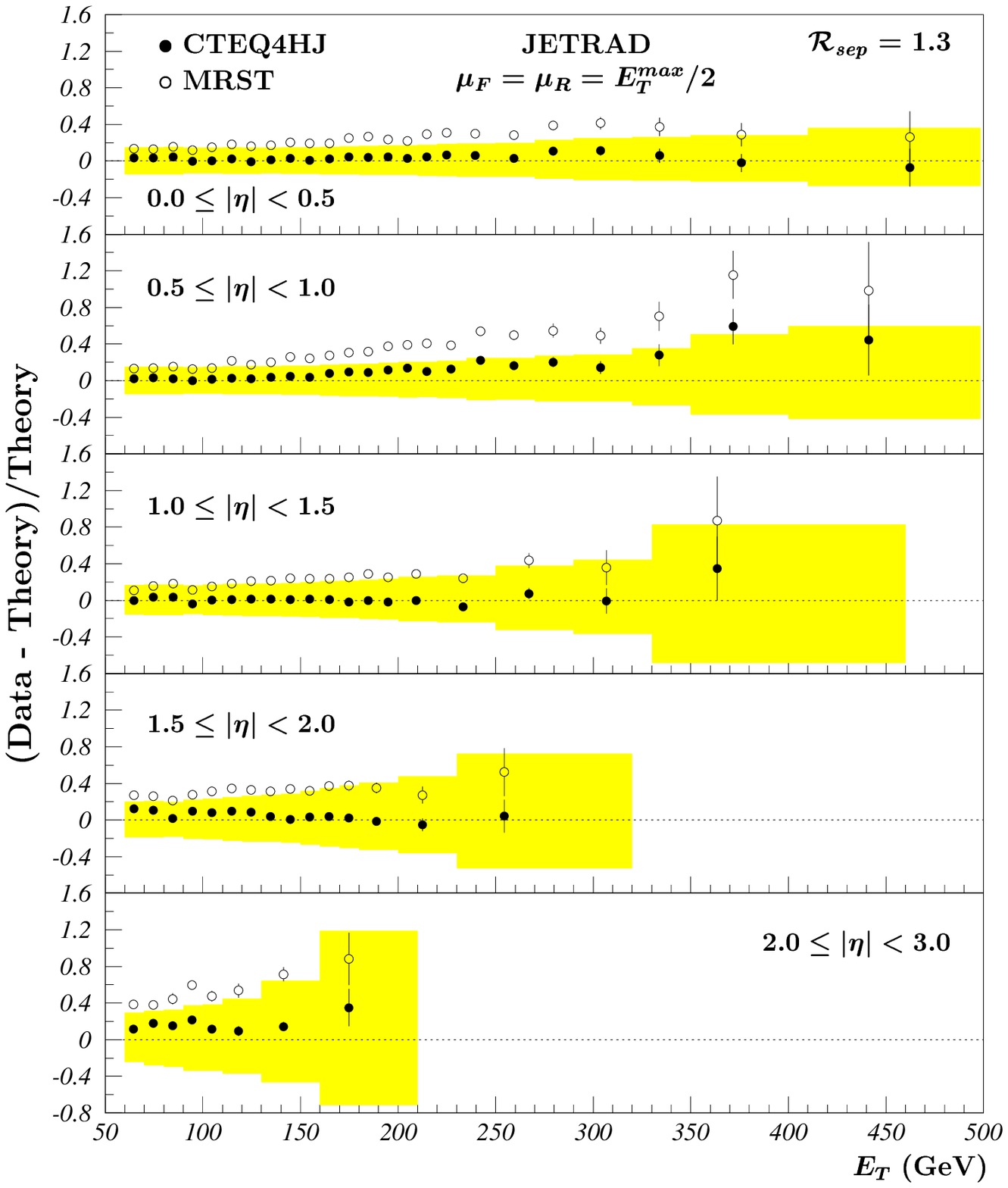,width=3.6in}}
\vskip-3cm  
\hskip1cm
\caption{Pseudorapidity dependence of the inclusive jet cross section 
($|\eta|<$3), compared with NLO QCD (CTEQ4HJ: full circles, MRTS: open 
circles). The bands represent the total systematic uncertainties in the
experiment.}
  \label{fig:forward}
\end{minipage}
\end{figure}

D\O\ has performed preliminary measurements of forward jet cross sections
up to pseudorapidities of $|\eta|=3$. These measurements allow to reach
regions in (x,$Q^{2}$) space previously unexplored. 
Figure~\ref{fig:forward} shows a comparison of the measured
cross sections in five different $\eta$ bins with the {\sc jetrad} 
prediction (CTEQ4HJ or MRST, and $\mu=E_{Tmax}/2$). 
Within the experimental and
theoretical uncertainties, the measurement and the calculation are in
good agreement. $\chi^{2}$ studies are underway for a quantitative statement.

\section{The Ratio of Dijet Mass Spectrums at  
$\sqrt{s}$~=~1800~GeV }

 The dijet mass spectrum is calculated using the relation:
 
 \begin{equation}
 \kappa \equiv \frac{d^{3} \sigma}{ d M_{JJ} d \eta_{1} d \eta_{2}} = 
 \frac { N_{i} }
  {{\cal L}_{i}  \epsilon_{i} \Delta  M_{JJ} \Delta \eta_{1}  
 \Delta \eta_{2}},
 \label{EQ:dijet_cross_section}
 \end{equation}
 
 where $N_i$ is the number of events in mass bin $i$; ${\cal L}_{i}$ is the integrated luminosity;
 $\epsilon_{i}$ is the efficiency of the trigger, vertex selection,
 and the jet quality cuts; $\Delta M_{JJ}$ is the width of the mass
 bin; and $\Delta \eta_{1,2}$ are the widths of the pseudorapidity bin
 At D\O\ the cross section is measured for the pseudorapidity bin
 $|\eta_{jet}| < 1.0$ . The systematic errors
 are dominated by the uncertainties due to the jet energy scale, which
 are 7$\%$ (30$\%$) for the 209 (873)~GeV mass bins. 

 The dijet mass cross section measurement was then repeated for
 $|\eta_{jet}| < 0.5$, and $0.5 < |\eta_{jet}| < 1.0$ and their ratio
 was determined. A large fraction of the total error cancels, as well
 as the uncertainty in the theoretical prediction
 of the ratio which is less than $3\%$ due to the choice of pdf, 
 and $6\%$
 from the choice of renormalization and factorization scale (excluding
 $\mu = 0.25 E_{Tmax}$). By taking the ratio
 $\kappa\left(|\eta_{jet}| < 0.5\right) / 
 \kappa\left(0.5 < |\eta_{jet}| < 1.0\right)$
 the systematic uncertainties decrease to less than 10$\%$.  

\begin{figure}[hbtp]
\vskip-1cm
\begin{minipage}[t]{3.05in}
\centerline{\psfig{figure=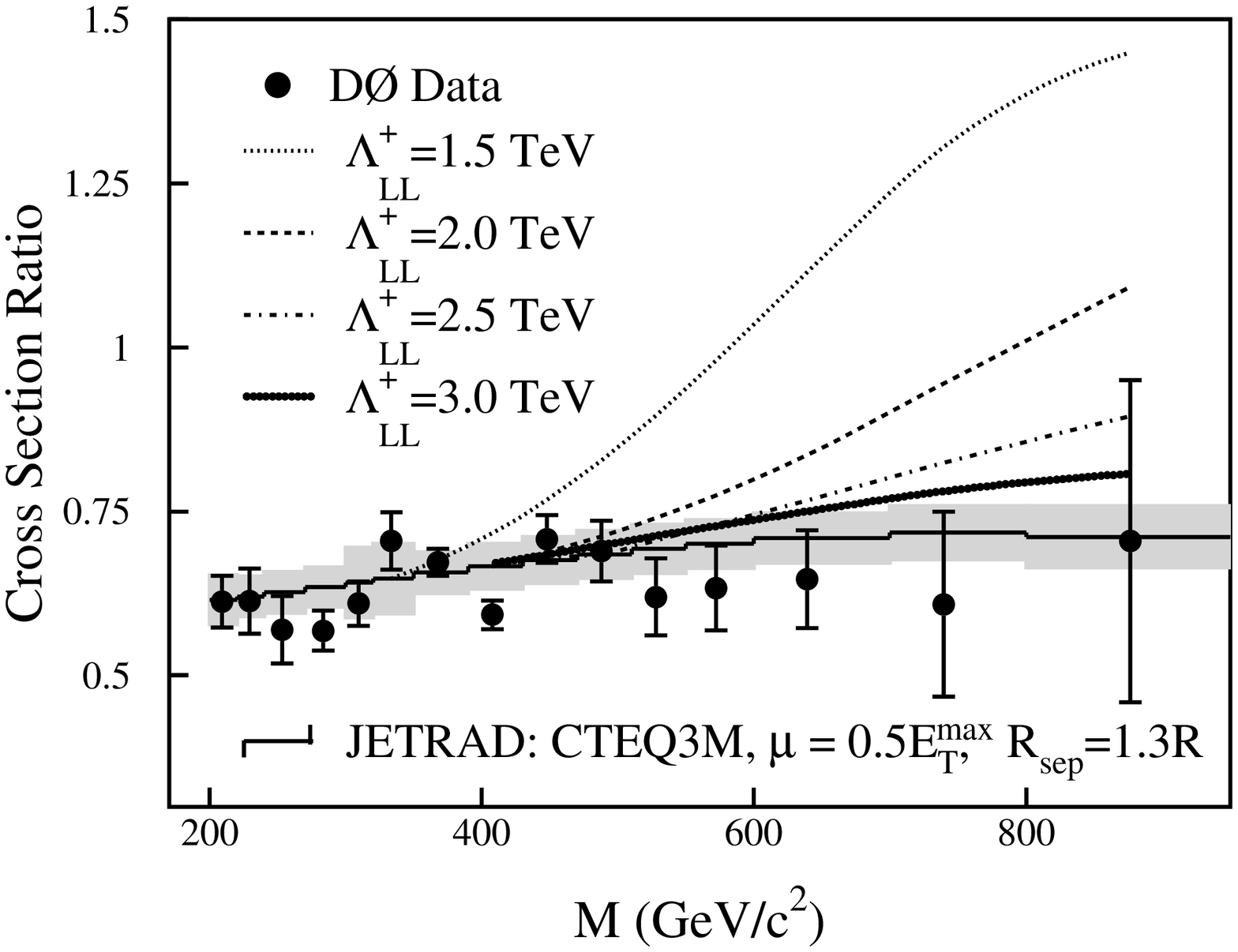,width=3.0in}}
\vskip-1cm 
\caption{The ratio of dijet mass cross sections for $|\eta_{jet}|<0.5$ and
 $0.5<|\eta_{jet}| <1.0$ for data (solid circles) and theoretical
 predictions for compositeness models with various values of
 $\Lambda_{LL}^{+}$. The error
 bars show the statistical uncertainties.  The shaded region
 represents the $\pm1\sigma$ systematic uncertainty.}
\label{fig:ratio_comp}
\end{minipage}
\end{figure}

 Given the experimental and theoretical
 uncertainties, the prediction can be regarded as in good agreement
 with the data (see Fig.~\ref{fig:ratio_comp}). The data are also in 
 agreement, within the
 uncertainties, with the cross section measured by
 \mbox{CDF}~\cite{dicdf}. 

 All choices of pdfs and renormalization scales 
 are in good agreement with the data ($\chi^{2}$) test, except for
 $\mu = 0.25 E_{Tmax}$ which is excluded by the data.

 The ratio of the mass spectra is used to place limits on quark
 compositeness. The
 {\sc pythia} event generator is used to simulate the effect of
 compositeness by taking the ratio of these LO predictions with 
 compositeness,
 to the LO with no compositeness, and scaling with this factor
 the {\sc jetrad}
 NLO prediction (shown in Fig.~\ref{fig:ratio_comp}).

 The D\O\ data shows no evidence of compositeness. The dijet mass spectrum rules 
 out
 quark compositeness models at the 95$\%$ confidence level where
 $\Lambda_{LL}^{+}$ is below 2.7~TeV and $\Lambda_{LL}^{-}$ is below
 2.4~TeV.
 
\section{The Triple Differential Jet Cross Sections at  
$\sqrt{s}$~=~1800~GeV }

Both CDF and D\O\ have performed preliminary measurements of triple
differential jet cross sections. This observable is defined only in
terms of the two leading jets of the event, to
provide information on pdfs. In particular, the Tevatron probes
high $Q^{2}$ and $x$ values previously unreachable.

CDF defines the triple differential jet cross section as 
$\frac{d^{3}\sigma}{dE_{T1}d\eta_{1} d\eta{2}}$, where $E_{T1}$ is the
transverse energy of a central jet in the event,
and $\eta_{1}$, $\eta{2}$ are the
pseudorapidities of the central and forward jets, respectively. 
This quantity is measured as a function of the transverse energy of 
the central jet for different $\eta_{2}$ bins up to $|\eta|$=3.

D\O\ measures $\frac{d^{3}\sigma}{dE_{T}d\eta_{1} d\eta{2}}$ versus $E_{T}$,
where $E_{T}$ is the transverse energy of the central or the forward jet
(the event enters twice in the measurement). This quantity is measured for 
different $\eta_{2}$ bins up to $|\eta|$=2. The main difference between the
CDF and D\O\ observables is that D\O\ measures the $E_{T}$ of both jets and
CDF measures only the $E_{T}$ of the central jet.

\begin{figure}[t]
\vskip-1cm
\begin{minipage}[t]{3.05in}
\centerline{\psfig{figure=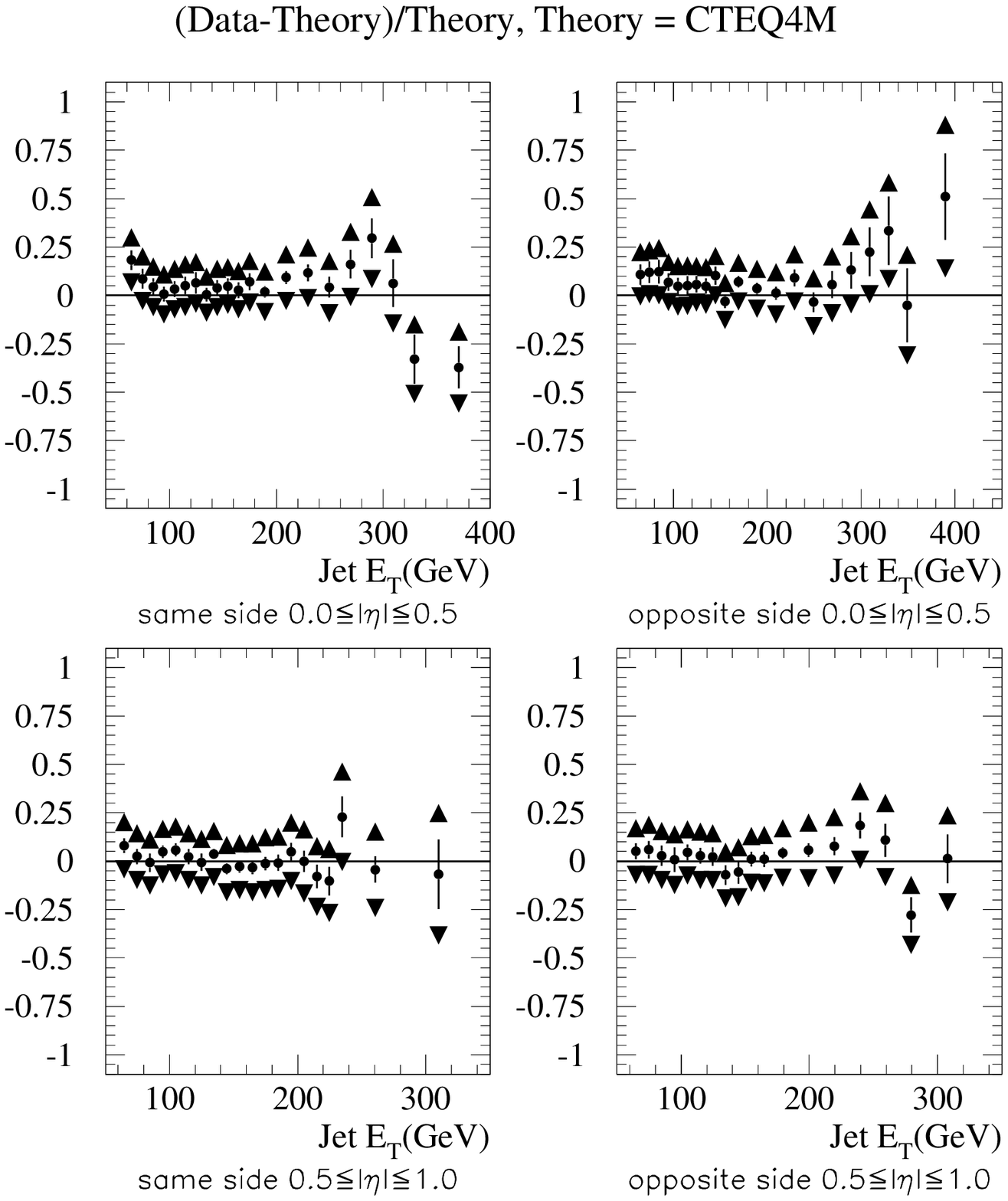,width=3.05in}}
\vskip-1.1cm  
\caption{D\O\ triple differential jet cross sections in different
 $|\eta_{2}|<2$ intervals for the
two leading jets (central and forward) in the same pseudorapidity side
(sign of $\eta_{1}$ same as sign of $\eta_{2}$).}
  \label{fig:freedy1}
\end{minipage}
\vskip-0.5cm
\hspace*{2mm}
\end{figure}

\begin{figure}[t]
\vskip-1.1cm  
\begin{minipage}[t]{3.05in}
\centerline{\psfig{figure=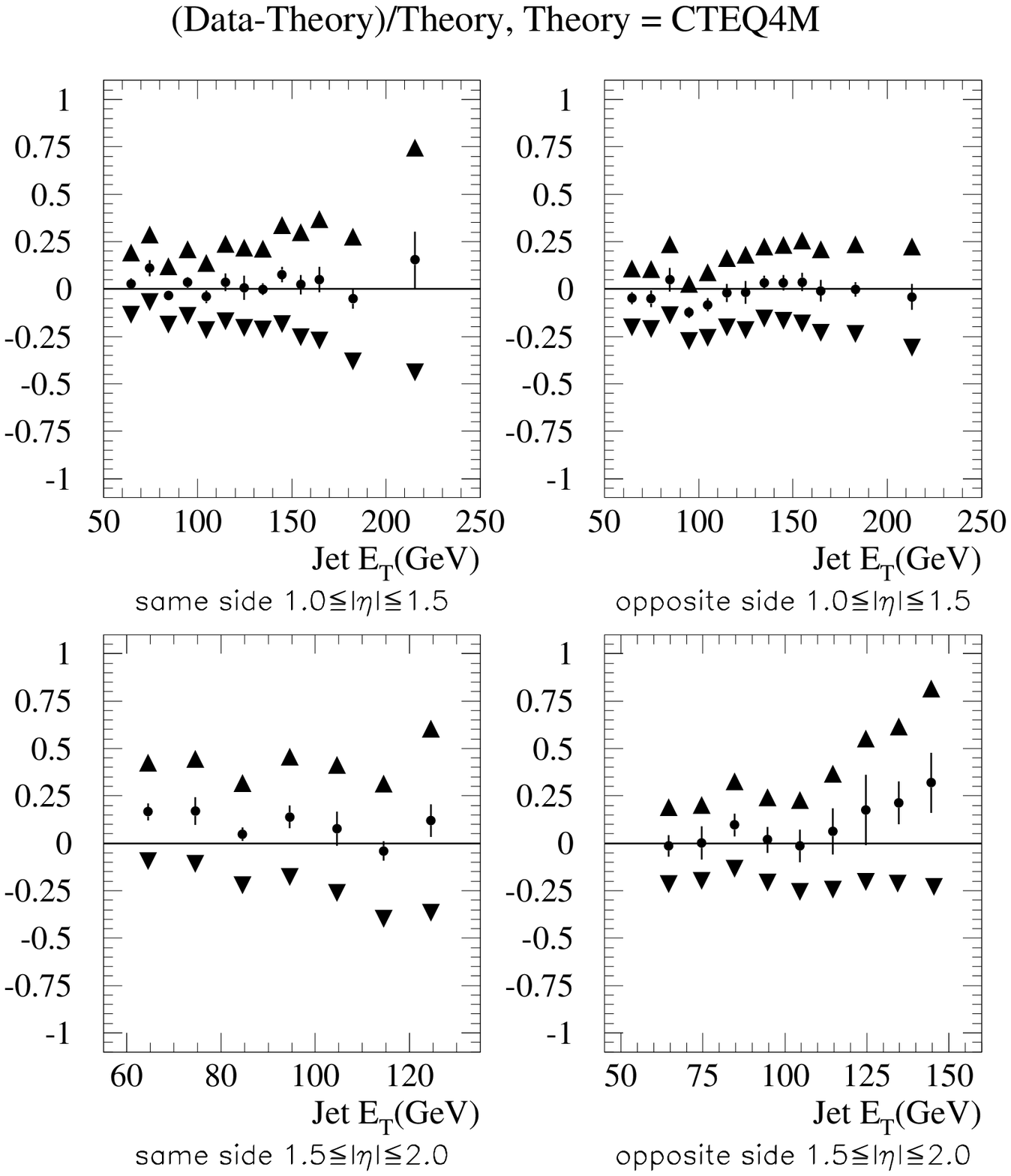,width=3.05in}}
\vskip-1cm
\caption{D\O\ triple differential jet cross sections in different
$|\eta_{2}|<2$ intervals for the
two leading jets (central and forward) in opposite pseudorapidity sides
(sign of $\eta_{1}$ opposite of sign of $\eta_{2}$).}
  \label{fig:freedy2}
\vskip1cm
\end{minipage}
\vskip-0.5cm
\hspace*{2mm}
\end{figure}

\begin{figure}[t]
\vskip-2.2cm
\begin{minipage}[t]{3.05in}
\centerline{\psfig{figure=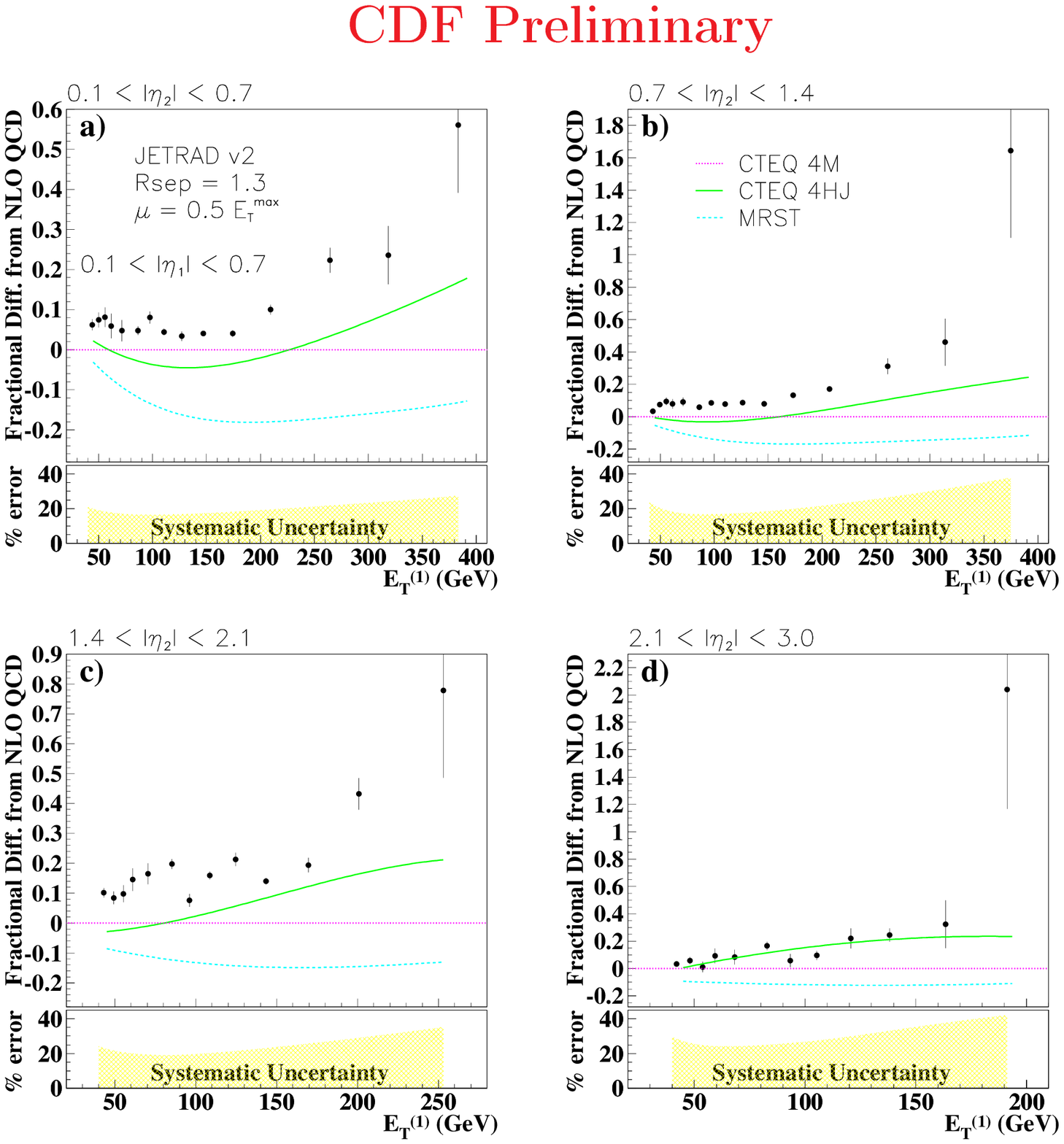,width=3.05in}}
\vskip-3.0cm  
\caption{CDF triple differential jet cross sections for different
$\eta_{2}$ bins.}
\vskip-1.5cm
  \label{fig:tripple_cdf}
\end{minipage}
\vskip-2cm
\hspace*{2mm}
\end{figure}

Figures~\ref{fig:freedy1}-~\ref{fig:freedy2} show the D\O\ measurement in different
$\eta_{2}$ bins for the
two leading jets (central and forward) in the same pseudorapidity side
(sign of $\eta_{1}$ same as sign of $\eta_{2}$), and opposite sides. Each
configuration is adequate for learning about different pdfs in different
regions of $(x,Q^{2})$ space. For example, two forward jets in the same
side are associated with one incoming parton with high x and the other
with low x.
Figure~\ref{fig:tripple_cdf} shows the CDF result for different
$\eta_{2}$ bins, independently of the relative sign of  $\eta_{1}$ and
$\eta_{2}$.  

Qualitatively, there is good agreement between data and theory. Both
experiments are currently working on quantitative studies.

\section{Dijet Cross Sections at Large $\eta$ Intervals}

At high center-of-mass energies, $\sqrt{s}$,
and for momentum transfers, $Q$, fixed and $\ll \sqrt{s}$, 
the radiative corrections to the parton-parton scattering contain
large logarithms $\ln(s/Q^2)$, which need to be summed to \mbox{all} 
\mbox{orders} in $\alpha_s$.
This summation is accomplished by the
Balitsky-Fadin-Kuraev-Lipatov (BFKL) equation~\cite{BFKL}.

Inclusive
dijet production provides an ideal possible signature of
BFKL dynamics.
For large values of the jet longitudinal momentum fraction, $x_j$,
the large logarithms $\ln(s/Q^2)$ result in large $\ln(\hat{s}/Q^2)$
(where $\sqrt{\hat{s}}$ is the partonic center-of-mass energy)
which factorize in the partonic dijet cross section, $\hat{\sigma}$.
The  $\ln(\hat{s}/Q^2)$ terms
are of the order of the pseudorapidity interval, $\Delta \eta$,
between the two jets~\cite{MN}
($\eta = -\ln(\tan(\theta/2))$, 
 where $\theta$ is the polar angle of the jet relative to 
 the proton beam).

D\O\ performed a measurement of the dijet cross section
at two different center-of-mass energies,
$\sqrt{s_A}=1800$~GeV and $\sqrt{s_B}=630$~GeV,
using the D\O\ detector at the Fermilab Tevatron.
The kinematics of the event is reconstructed using the most 
forward/backward jets, and
the cross section is measured as a function of $x_1$, $x_2$ and $Q^2$
at each center-of-mass energy.
The ratio of the cross sections is then determined
at the same values of $x_1$, $x_2$ and $Q^2$ between the two energies.
This eliminates the dependence of the cross section on the pdf's 
and reduces the ratio to that of the partonic cross sections.
It can be shown~\cite{bfkld0} that the latter is a function only of 
the pseudorapidity separations between the two jets ($\Delta \eta$'s):
\begin{equation}
R = \frac{\hat{\sigma}(\Delta \eta_A)}
         {\hat{\sigma}(\Delta \eta_B)}
  = \frac{e^{(\alpha_{BFKL}-1)(\Delta \eta_A-\Delta \eta_B)}}{\sqrt{\Delta \eta_A/\Delta \eta_B}} \;.
\label{eq:ratio}
\end{equation}
In other words, variation of $\sqrt{s}$,
while keeping $x_1$, $x_2$ and $Q^2$ fixed,
is equivalent to variation of $\Delta \eta$,
which directly probes the BFKL dynamics.

Several theoretical predictions can be compared to the D\O\ measurement.
Leading Order QCD predicts the ratio of the cross sections
to fall asymptotically toward unity.
The {\sc herwig}~\cite{herwig}
 Monte Carlo provides a more realistic prediction.
It calculates the exact $2 \rightarrow 2$ subprocess including
initial and final state radiation and angular ordering of the emitted 
partons.
The LLA BFKL intercept for
$\alpha_s(20\,{\rm GeV})=0.17$~\cite{Orr-Stirling} is equal to 1.45.
The Next-to-Leading Logarithmic~\cite{BFKL-NLL}
are not as yet available.

\begin{figure}[htb]
\vskip-1.2cm
\begin{minipage}[t]{3.05in}
\centerline{\psfig{figure=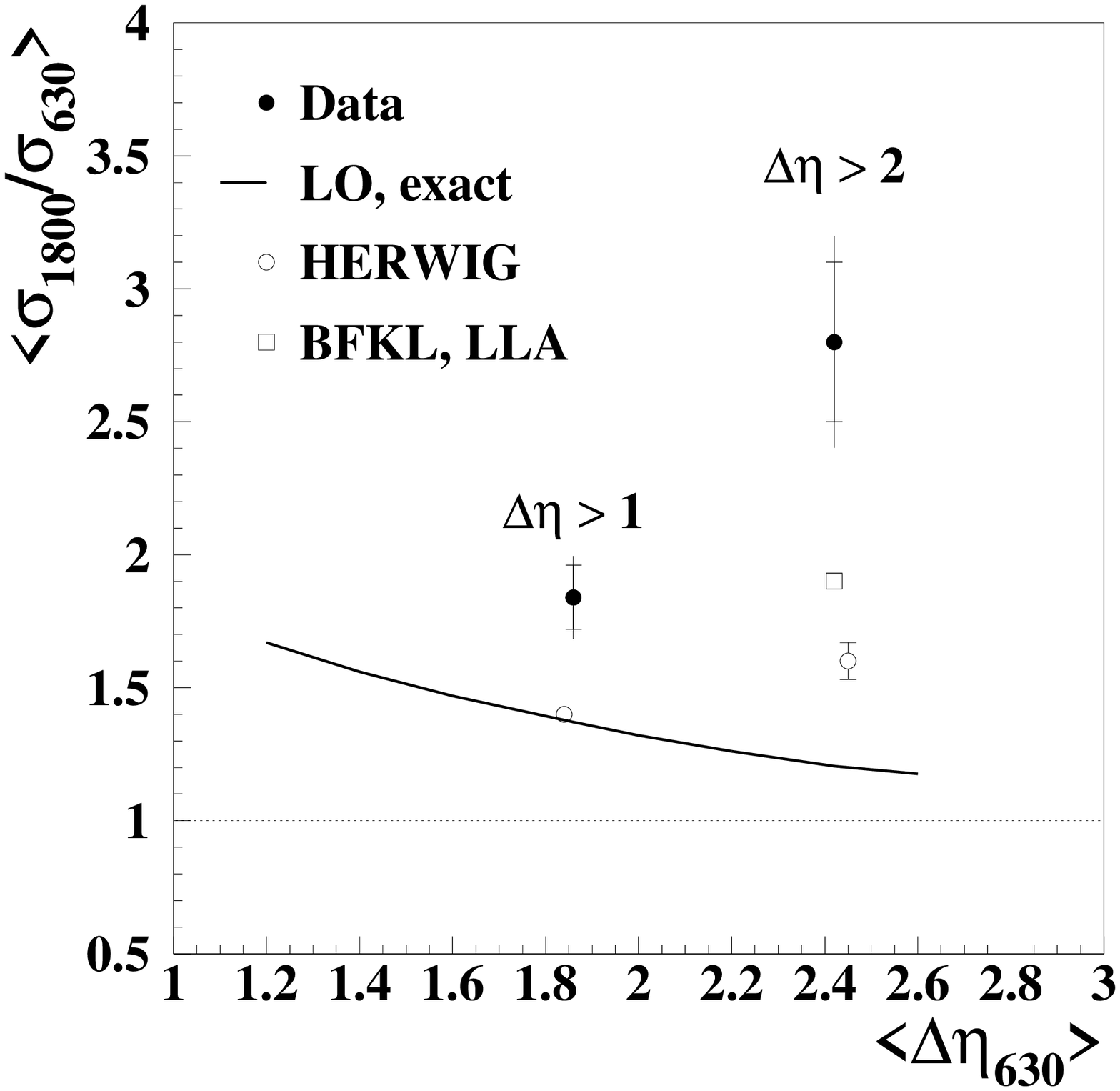,width=3.05in}}
\vskip-1.1cm  
\caption{The D\O\ ratio of the dijet cross sections at 
         both center-of-mass energies
         for $\Delta \eta>1$ and $\Delta \eta>2$.
         The inner error bars on the data points
         represent statistical uncertainties;
         the outer bars represent statistical and uncorrelated
         systematic uncertainties added in quadrature.
         The error bars on the {\sc herwig} predictions
         represent statistical uncertainties.}
  \label{fig:ratio_vs_y}
\end{minipage}
\vskip-0.5cm
\end{figure}

The ratio of cross sections is shown in Fig.~\ref{fig:ratio_vs_y}
as a function of the mean pseudorapidity interval at 630~GeV.
It is evident that the growth of the dijet cross section with 
 $\Delta \eta$
is stronger in the data than in any theoretical model which was
considered.
Namely, the measured ratio is higher by 4 standard deviations
than the LO prediction, 3 deviations than the {\sc herwig} 
prediction, and 2.3 deviations than the LLA BFKL one.

\section{Subjet Multiplicities}

A jet is typically associated with the energy and momentum of each 
final state parton. Experimentally, however, it is a cluster of energy
in the calorimeter. 
QCD predicts that gluons radiate more than quarks.  
Asymptotically, the ratio of objects within gluon jets
to quark jets is expected to be in the ratio of their color charges 
$C_A / C_F = 9 / 4$\cite{Ellis}.

D\O\ performed a preliminary measurement of subjet multiplicities in 
quark 
and gluon jets, as well as the ratio of the means of these two 
quantities.
For this analysis, jets are reconstructed using the $K_{T}$ 
algorithm~\cite{ES,cat92,cat93} with a resolution parameter
$\cal{D}$=1 (see Ref.~\cite{rob99}).

$M$ is the subjet multiplicity 
in a mixed sample of quark and gluon jets.
It may be written as a linear combination of subjet multiplicity
in gluon and quark jets:

\begin{equation}
M=fM_{g}+(1-f)M_{q}
\label{eq:m}
\end{equation}

The coefficients are the fractions of gluon and
quark jets in the sample, $f$ and $(1-f)$, respectively. 
Consider Eq. (\ref{eq:m}) for two
samples  of jets in the same kinematic range, one
at $\sqrt{s} = 1800$ (gluon dominated) 
and the other at 630~GeV (quark dominated),
assuming $M_{g}$ and $M_{q}$
are independent of $\sqrt{s}$.
The solutions are

\begin{equation}
M_{q}=\frac{f^{1800}M^{630}-f^{630}M^{1800}}{f^{1800}-f^{630}} 
 \label{eq:mq}
\end{equation}

\begin{equation}
M_{g}=\frac{\left( 1-f^{630}\right) M^{1800}-\left( 1-f^{1800}\right) M^{630}}
      {f^{1800}-f^{630}} 
\label{eq:mg}
\end{equation}

where $M^{1800}$ and $M^{630}$ are the
experimental measurements in the mixed jet samples at  
$\sqrt{s} = 1800$ and 630~GeV,
and $f^{1800}$ and $f^{630}$ are the gluon jet fractions in the two samples.
The method relies on knowledge of the two gluon jet fractions.

\begin{figure}[htb]
\vskip-1.5cm
\begin{minipage}[t]{3.05in}
\centerline{\psfig{figure=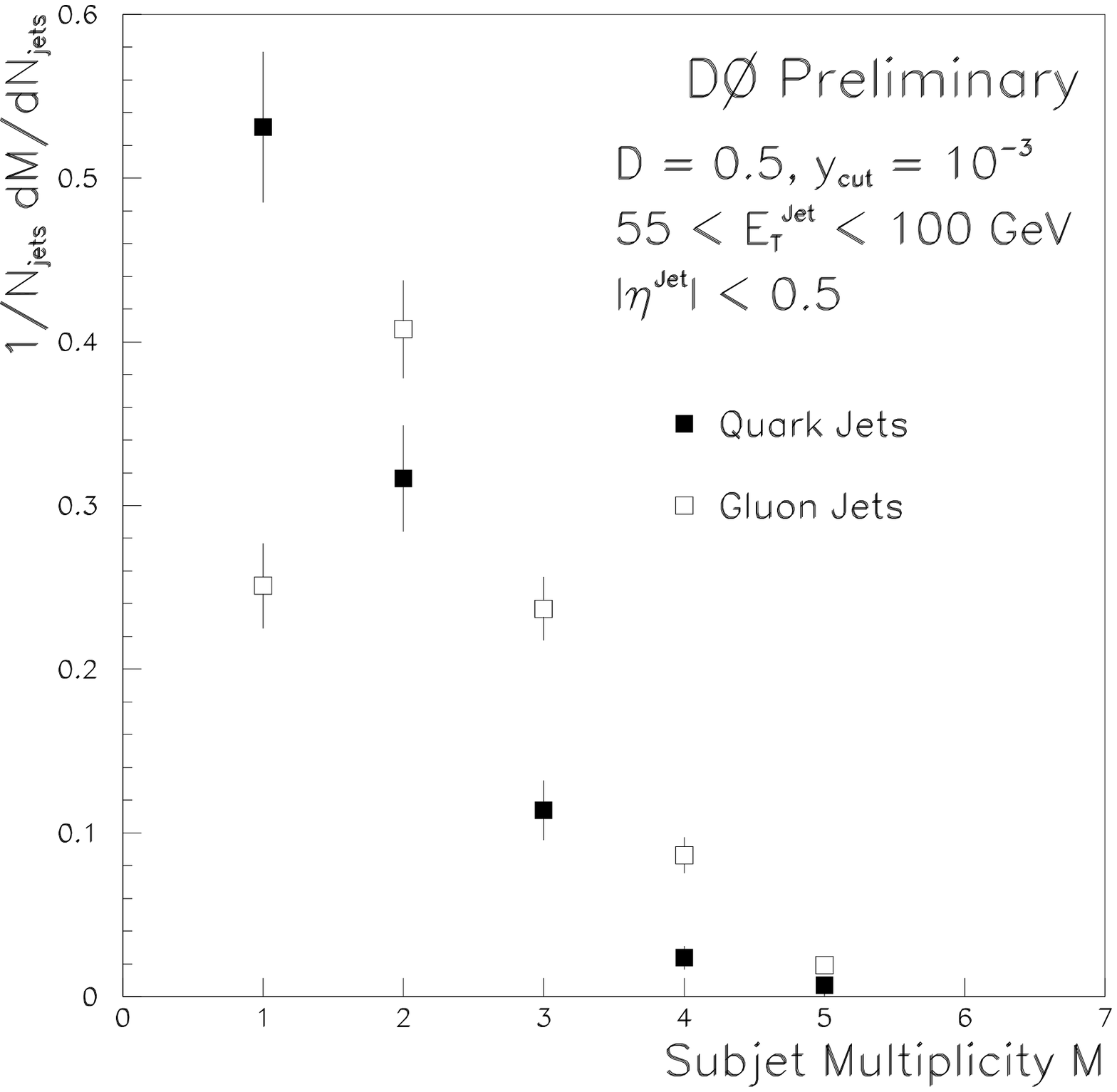,height=3.05in}}
\vskip-1cm
\caption{Corrected subjet multiplicity in quark and gluon jets, 
extracted from D\O\ data.}
\label{fig:qg}
\end{minipage}
\vskip-1cm
\end{figure}

Figure~\ref{fig:qg} shows that the subjet multiplicity is 
clearly larger for gluon jets
compared to quark jets.
The gluon jet fractions are the largest
source of systematic error. 
The measured ratio and its total uncertainty are:
$R = \frac{\langle M_g \rangle - 1} {\langle M_q \rangle - 1}
= 1.91 \pm 0.04{\rm(stat)} ^{+0.23}_{-0.19} {\rm(sys)}$.
The ratio is well described by the {\small HERWIG} parton 
shower Monte Carlo, and is only slightly 
smaller than the naive QCD prediction 9/4.

\section{Particle Multiplicities}

Perturbative QCD calculations, carried out in the framework of the 
Modified Leading Log Approximation~\cite{mlla} (MLLA), complemented 
with the Local Parton-Hadron Duality Hypothesis~\cite{lphd}  (LPHD), 
predict the shape of the momentum distribution, as well as the total 
inclusive multiplicity, of particles in jets. The MLLA is an
asymptotic calculation, which proves to be infrared stable, in the 
sense that the model cutoff parameter $Q_{eff}$ can be safely pushed 
down to $\Lambda_{QCD}$. LPHD is responsible for the hadronization stage 
and implies that hadronization is local and happens at the end of the
parton shower development. In its simplest interpretation, the model 
has one parameter $K_{LPHD}$, the rate of parton-to-hadron conversion: 
\begin{equation} 
N_{hadrons} = K_{LPHD} \times N_{partons}. 
\label{eq:murnf} 
\end{equation} 

In MLLA, momentum distributions and multiplicities in quark and gluon 
jets in a restricted cone of size $\theta$ around the jet axis are 
functions of $E_{jet}\theta/Q_{eff}$~\cite{formulas} and differ by a 
factor $r$:

\[
N^{q-jet}  ( \xi )  = \frac{1}{r} N^{g-jet}( \xi )
\]

\begin{equation}
\xi = \log{\frac{1}{x}}, x=p_{track}/E_{jet}
\label{eq:mu2}
\end{equation}

Jets at the Tevatron are a mixture of quark and gluon jets. Therefore,

\[
N^{charged}_{hadrons} ( \xi )  = 
\]

\[
K^{charged}_{LPHD} ( \epsilon_{g}+(1-\epsilon_{g})\frac{1}{r})
F^{nMLLA} N^{q-jet}_{part}( \xi ) =
\]

\begin{equation}
 K  N^{q-jet}_{part}( \xi )  
\label{eq:mu3}
\end{equation}

where $\epsilon_g$ is the fraction of gluon jets in the events, 
the factor of $1/r$ reflects the difference between gluon and 
quark jets, and, finally, the factor $F^{nMLLA}$ accounts for the 
next-to MLLA corrections to the gluon spectrum. Theoretical 
calculations~\cite{theor1} predict somewhat different values of 
$F^{nMLLA}$, but all agree that $F^{nMLLA}$ has almost no dependence 
on the jet energy in the region relevant to this analysis.  
The average of the results above was chosen and the difference 
between predictions was used as a theoretical error: 
$F^{nMLLA}$=1.3$\pm$0.2. 
The same papers predict the value of $r$ to be between 1.5 and 1.8.

\subsection{The Dijet Data Analysis}

CDF data collected during the 1993-1995 running period was used for
this analysis. Events with two jets well balanced in 
transverse energy were selected. Both jets were required to be in the 
central region. Tracks were counted 
in restricted cones of sizes 0.28,0.36 and 0.47 around the jet axis.

\begin{figure}[t]
\begin{minipage}[t]{3.05in}
\centerline{\psfig{figure=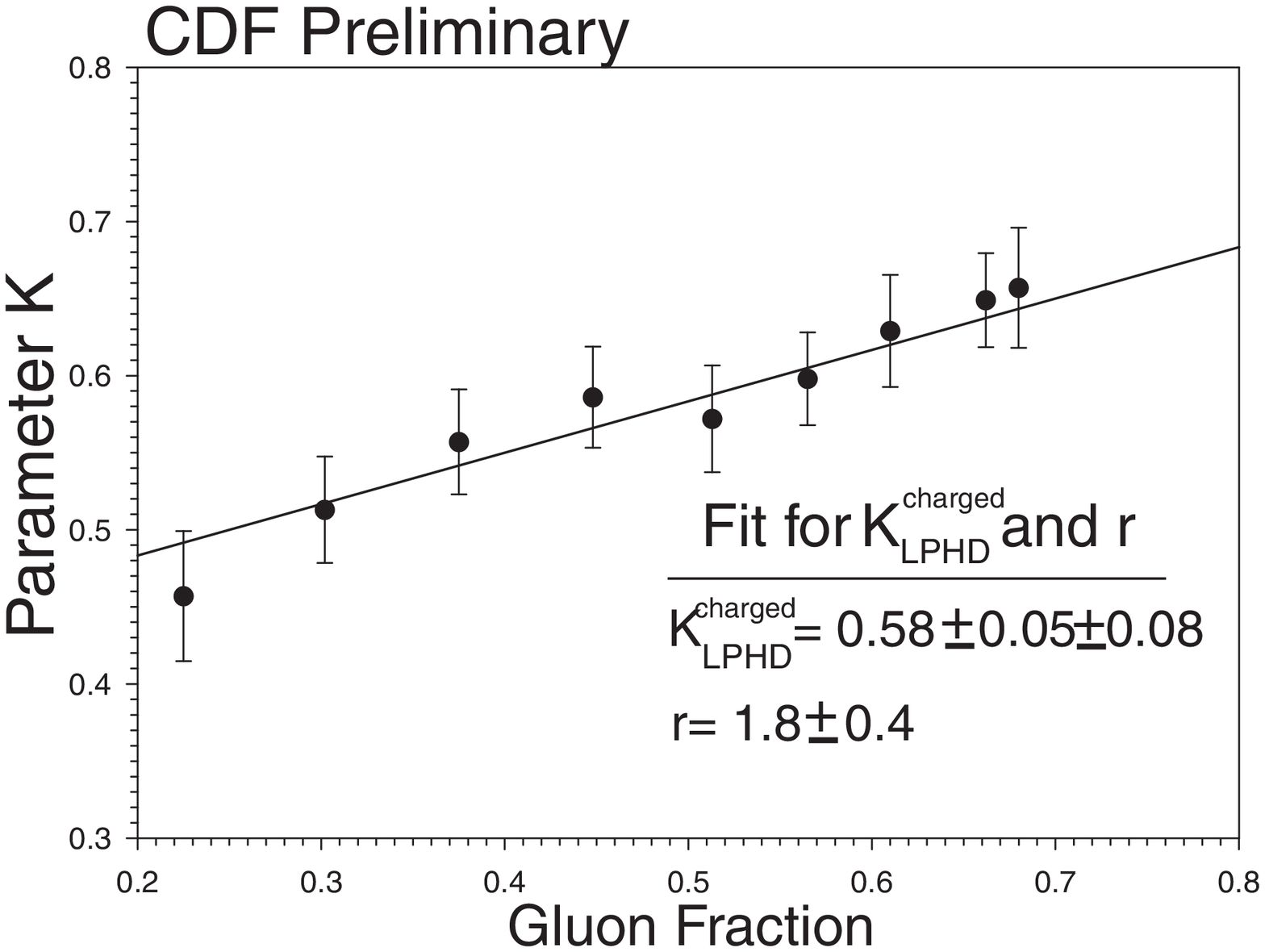,height=2.1in}}
\vskip-1cm
\caption{Fit of the parameter $K$ 
for $K_{LPHD}$ and $r$. Cone 0.47. First error is combined statistical
and systematic errors, 
the second one - theoretical error coming from $F^{nMLLA}$.}  
\label{fig:kvsgluon}
\vskip-1cm
\end{minipage}
\end{figure}

\begin{figure}[t]
\begin{minipage}[t]{3.05in}
\centerline{\psfig{figure=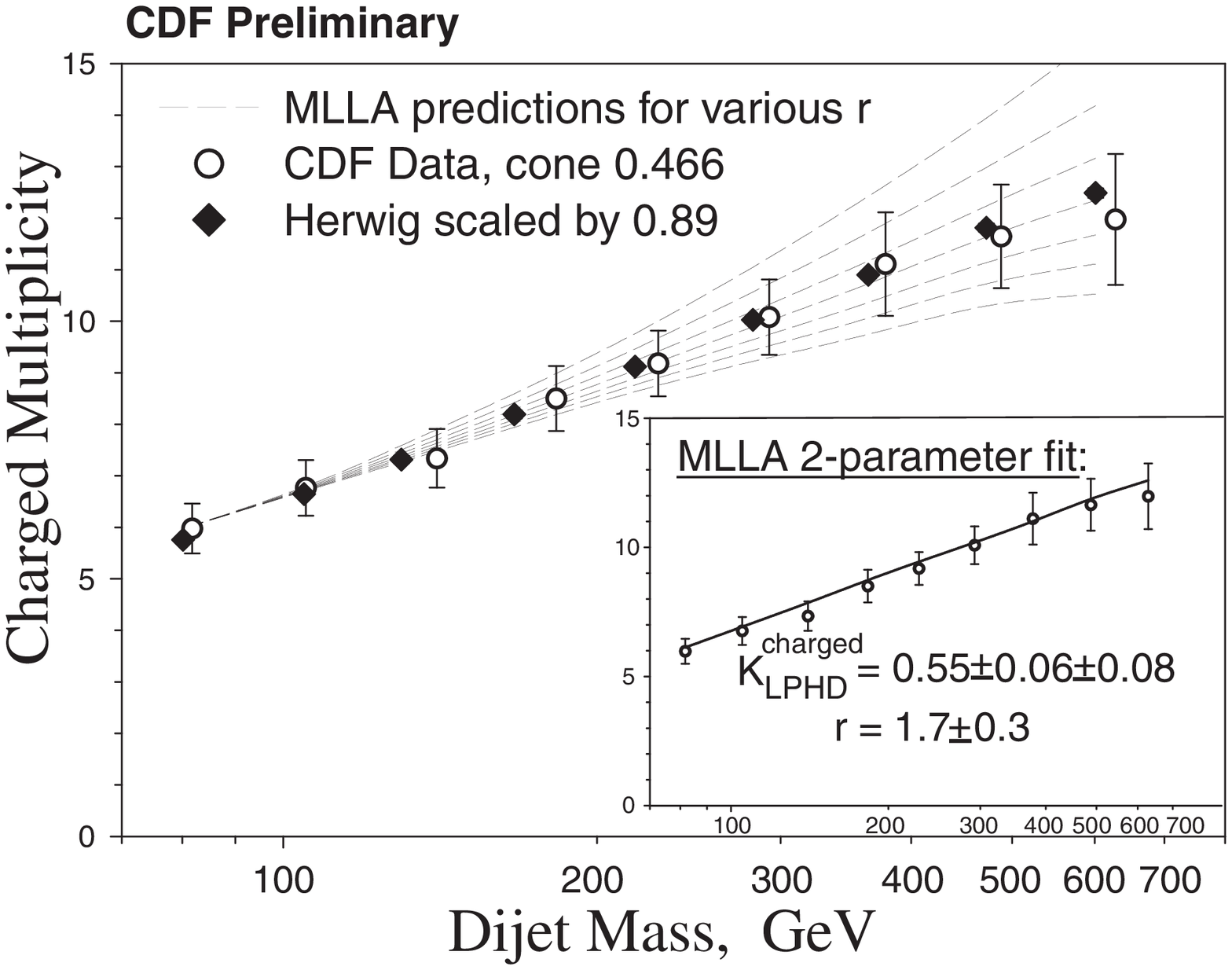,height=2.1in}}
\vskip-1cm
\caption{Charged particle multiplicity (per jet) as a 
function of the dijet mass. MLLA fit for
$K_{LPHD}$
and $r$.}
\vskip-1cm 
\label{fig:mult1}
\end{minipage}
\end{figure}

Analysis of the fitted parameter K allows an extraction of both 
$K_{LPHD}$ and r. According to Eq.~\ref{eq:mu3}, the dependence is linear. 
Figure~\ref{fig:kvsgluon}, 
shows 9 values of K (corresponding to 9 dijet masses for the 
largest cone-size 0.47) vs the gluon jet fraction (extracted 
using Herwig 5.6) in the events from respective dijet mass bins, 
as well as the results of the fit for $K_{LPHD}$ and $r$.
The same parameters can be extracted from the inclusive 
multiplicity using an integrated version of
Eq.~\ref{eq:mu3}. In this case, the extracted parameters will only rely on the 
total multiplicity and not on
the exact shape of the distribution. Figure~\ref{fig:mult1}
shows the fit of data with MLLA predictions as well as
the fitted parameters $K_{LPHD}$ and $r$. It is remarkable that the 
two results are in such a good agreement.

\subsection{Model-Independent Measurement}

\begin{figure}[t]
\begin{minipage}[t]{3.05in}
\centerline{\psfig{figure=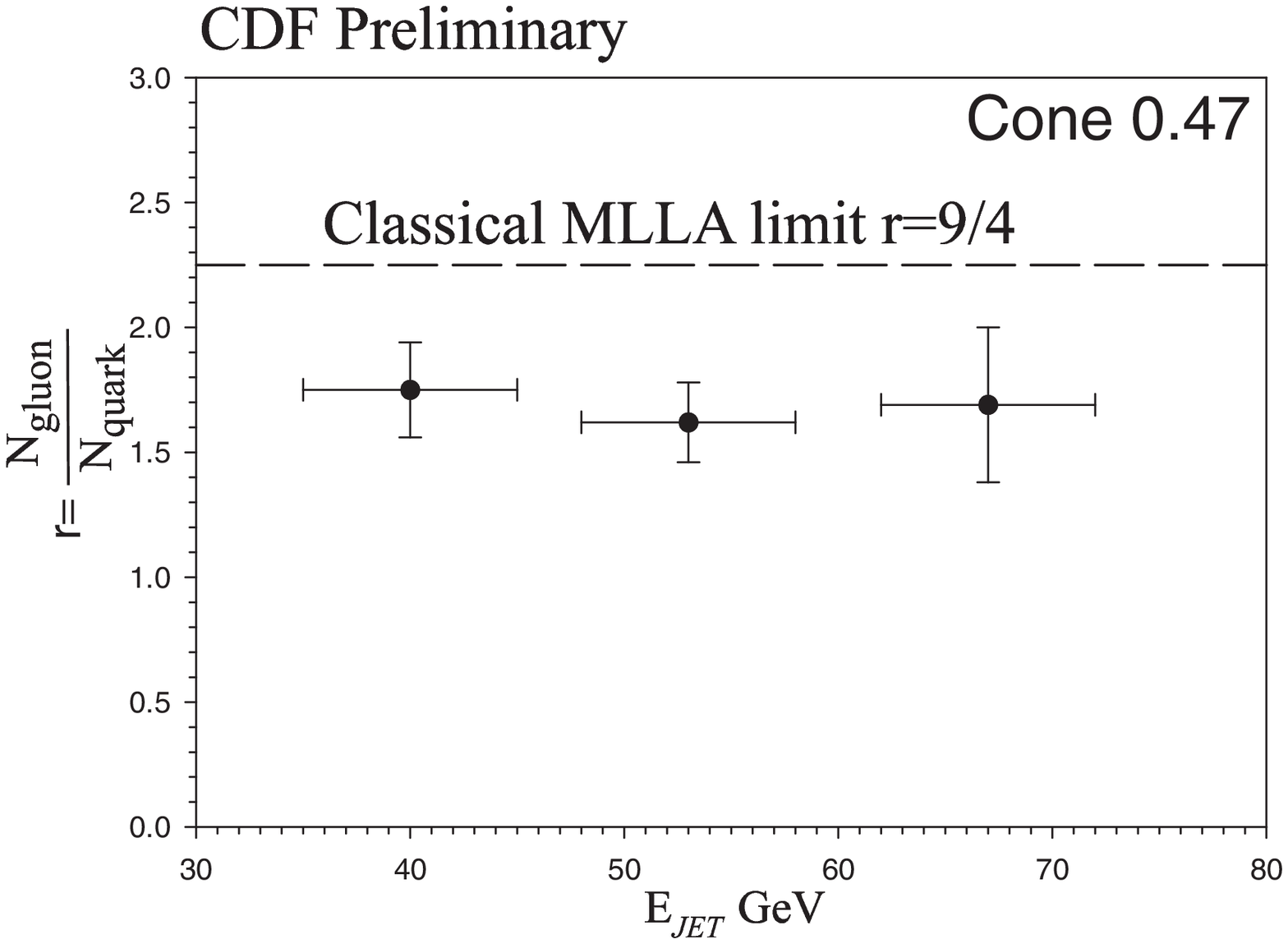,height=2.1in}}
\vskip-1cm
\caption{Ratio of charged multiplicities in gluon and quark jets 
based on comparison of the dijet and
$\gamma$-jet events.}  
\vskip-0.5cm 
\label{fig:mult2}
\end{minipage}
\end{figure}

The multiplicity in dijet and $\gamma$-jet events is compared 
(data selection was similar) to extract model-independent 
measurement of r. These samples have very different fraction of 
gluon jets for the jet energies 40-60~GeV (roughly 60\% for dijets and
12\% for $\gamma$-jet, according to Herwig 5.6). The multiplicities 
measured for each of the samples and a
knowledge of the gluon jet fractions allowed to extract $r$. 
Figure~\ref{fig:mult2} shows the measured r as a
function of the jet energy. The result for $r$ is 
1.75$\pm$0.11$\pm$0.15 
in perfect agreement with MLLA result.

\section{Conclusions}

The 1992-1996 collider run at Fermilab represented a major step
in the testing of QCD. The D\O\ and CDF experiments measured
jet cross sections with unprecedented accuracy, extended the
energy reach to $\sim$450~GeV, and set a new limit of
2.4-2.7~TeV for the quark compositeness scale. In general, and 
within experimental and
theoretical uncertainties, QCD is in good agreement with the data.
Measurements on jet
structure and fragmentation were also performed and yielded agreement
with QCD; they support the perturbative nature 
of jet fragmentation. The upcoming run at the Tevatron,
scheduled to start in March 2001, will extend the energy
frontier even further ($\sqrt{s}$=2~TeV) and collect at least
20 times more data, allowing precision measurements
of QCD in kinematic regions previously unexplored.

\end{document}